\documentclass[reprint,
 amsmath,amssymb,
 aps,
showkeys,
showpacs
]{revtex4-2}

\usepackage{comment}
\usepackage{graphicx}
\usepackage{dcolumn}
\usepackage{bm}

\begin{document}

\preprint{MUPB/Conference section: }

\title{Muon g-2 and other observables \\ in models with extended Higgs and matter sectors}
\thanks{Invited review talk at 20th Lomonosov Conference, Moscow, Russia, August 21, 2021.}%

\author{Radovan Dermisek}
 \affiliation{Physics Department, Indiana University, Bloomington, IN, 47405, USA}


\date{January 15, 2022}

\begin{abstract}
I review possible explanations of the muon g-2 anomaly in models with extended Higgs and matter sectors, focusing on  extensions of the standard model and the two Higgs doublet model with vectorlike leptons. 
Predictions of these models, namely the modifications of muon Yukawa and gauge couplings, that can be searched for at the LHC and future colliders, are summarized. I also discuss striking predictions for di-Higgs and tri-Higgs signals at a muon collider that can be tested even at very low energies. Furthermore, I briefly comment on other interesting features and signatures of models with extended Higgs sector and vectorlike matter.

\end{abstract}

\maketitle


\newpage 
\section{Introduction}\label{intro}

\begin{figure}[t]
\includegraphics[scale=0.5]{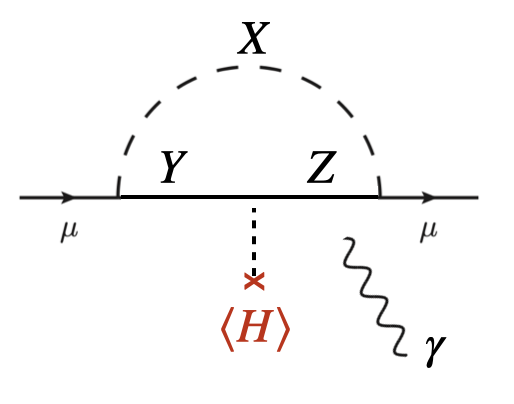}
\caption{A generic mass enhanced diagram contributing to $(g-2)_{\mu}$.
}
\label{fig:XYZ}
\end{figure}

An interesting thing about the deviation of the measured value of muon anomalous magnetic moment from the standard model (SM) prediction is that it can be comfortably explained by new particles with masses far above the reach of the Large Hadron Collider (LHC). The mass enhanced contributions of new particles to $a_\mu = (g-2)_{\mu}/2$, see Fig.~\ref{fig:XYZ}, scale as
\begin{equation}
\Delta a_{\mu} \simeq \frac{\lambda_{NP}^3}{16\pi^2}  \frac{m_{\mu}v}{m_{NP}^2}.
\label{eq:NP}
\end{equation}
Thus, the 4.2$\sigma$ anomaly, $\Delta a^{exp}_{\mu}\equiv a^{exp}_{\mu} - a_{\mu}^{SM} = (2.51 \pm 0.59)\times 10^{-9}$~\cite{Abi:2021gix, Aoyama:2020ynm}, points to a multi-TeV scale 
of new physics for  order one couplings, and it extends to about  $50$ TeV for couplings close to the perturbativity limit. 

In contrast, if the muon is in the internal fermion line, the contribution scales as $(\lambda_{NP}^2/{16\pi^2})  m_{\mu}^2/m_{NP}^2$, suggesting the electroweak scale as the scale of new physics for order one couplings. Picking up the vacuum expectation value (vev) by a new particle in the loop rather than by the muon leads to a dramatic $\lambda_{NP} v /m_\mu$ enhancement. 

Particles $X$, $Y$, and $Z$ in Fig.~\ref{fig:XYZ} can have any quantum numbers as long as the loop can be closed and the vev can be picked up on the way. This leads to a large number of possible models. For the same size of couplings and masses all the models generate similar contributions to $(g-2)_{\mu}$. However, the models highly vary with respect to what ranges of couplings and masses are allowed by other constraints specific to a given model. 

Perhaps the simplest and the most economical scenarios are those with just two new fields in the loop, namely that the role of $X$ in Fig.~\ref{fig:XYZ}  is played by the SM Higgs or gauge bosons, and $Y$ and $Z$ are  new vectorlike leptons~\cite{Kannike:2011ng, Dermisek:2013gta}. There are several choices for their quantum numbers, including the quantum numbers of the SM lepton doublet and charged singlet that we will focus on. These models are constrained by the modifications of muon Yukawa and gauge couplings that result from the mixing of the muon with new leptons. 

The least constrained and perhaps because of that the most popular scenarios seem to be models with three new fields in the loop: two vectorlike leptons and one new scalar that does not participate in electroweak symmetry breaking (EWSB) and thus does not result in the mixing of the muon with new leptons. There are numerous possibilities for their quantum numbers, for a summary see Ref.~\cite{Capdevilla:2020qel}. Such scenarios can explain the $(g-2)_{\mu}$ measurement with ease while leaving negligible traces at current (or even future) colliders. On the other hand, introducing three new fields that otherwise do nothing but have just the right quantum numbers to pick up the vev and close the loop seems somewhat too contrived. 

Somewhere in between the previous two scenarios are models with vectorlike leptons and new scalars that participate in EWSB and appear in well studied models. For example, in  two Higgs doublet models (2HDM) the role of $X$ in Fig.~\ref{fig:XYZ}  can be played by heavy Higgs bosons: $H$, $A$, and $H^\pm$, in addition to the SM Higgs and gauge bosons. Interestingly, the type-II 2HDM with vectorlike leptons interpolates between the SM case (for small $\tan\beta$) and the models with new scalars not participating in EWSB (large $\tan \beta$)~\cite{Dermisek:2020cod, Dermisek:2021ajd}. 

For other specific models involving vectorlike leptons and scalars, and many other possibilities for new particles in the loop in Fig.~\ref{fig:XYZ} that include examples with new gauge fields, new quarks and leptoquarks, and  models with superpartners, see for example Refs.~\cite{Megias:2017dzd, Choudhury:2017fuu, Kowalska:2017iqv, Raby:2017igl, Calibbi:2018rzv, Crivellin:2018qmi, Barman:2018jhz, Arnan:2019uhr, Kawamura:2019rth, Hiller:2019mou, Endo:2020tkb, Frank:2020smf, Chun:2020uzw, Kowalska:2020zve, Mondal:2021vou, Ferreira:2021gke, Crivellin:2021rbq, Escribano:2021css, Alvarado:2021nxy, Arkani-Hamed:2021xlp, Hue:2021zyw, Hue:2021xzl, Lee:2021gnw, He:2021yck}.

In what follows I will focus on the type-II 2HDM with vectorlike leptons as the main example, highlight interesting features of the SM case, and only briefly comment on some other scenarios. Before that, however, I will briefly mention few other theoretical and phenomenological aspects of models with extended Higgs and matter sectors.
 
Extensions of the Higgs sector of the SM, especially two Higgs doublet models, and the extensions of the matter sector by vectorlike pairs of quarks and leptons have been explored in a variety of contexts. 
Vectorlike quarks and leptons with analogous quantum numbers to SM quarks and leptons allow for a straightforward embedding into grand unified theories. They also allow for a novel understanding of values of gauge couplings and the third generation Yukawa couplings. 
Adding complete vectorlike familes to the SM provides a possible explanation for the observed hierarchy of gauge couplings~\cite{Dermisek:2012as,Dermisek:2012ke}. The minimal supersymmetric model  with a complete vectorlike family in the multi-TeV range can explain values of the seven largest couplings in the SM~\cite{Dermisek:2017ihj,Dermisek:2018hxq,Dermisek:2018ujw} from the IR fixed point structure of renormalization group equations. Especially the prediction for the value of the smallest gauge coupling, $\alpha_1$, is striking. Multi-TeV vectorlike quarks and squarks  can also remove the contribution to the Higgs mass from the RG flow above their scale in supersymmetric models~\cite{Dermisek:2016tzw}. Furthermore, vectorlike fermions were suggested  to explain various anomalies, for example discrepancies in precision Z-pole observables~\cite{Choudhury:2001hs, Dermisek:2011xu, Dermisek:2012qx, Batell:2012ca}.

\section{Muon g-2 in 2HDM with vectorlike leptons}\label{g-2}

The main results are presented for a 2HDM extended by vectorlike pairs of new leptons, $L_{L,R}$ and $E_{L,R}$, that mix with the second generation of SM leptons. The general Lagrangian describing masses and interactions in a type-II  version of the model, that can be enforced by a $Z_2$ symmetry, is given by~\cite{Dermisek:2020cod}:
\begin{flalign}
\mathcal{L}\supset& - y_{\mu}\bar{l}_L\mu_{R}H_{d} - \lambda_{E}\bar{l}_{L}E_{R}H_{d}  - \lambda_{L}\bar{L}_{L}\mu_{R}H_{d}   \nonumber \\   
& - \lambda\bar{L}_{L}E_{R}H_{d}   - \bar{\lambda}H_{d}^{\dagger}\bar{E}_{L}L_{R} \nonumber \\ 
	& - M_{L}\bar{L}_{L}L_{R} - M_{E}\bar{E}_{L}E_{R}  + h.c..
\label{eq:lagrangian}	
\end{flalign}
Let us label the doublet components as $l_{L}=( \nu_{\mu},  \mu_{L} )^T$,  $ L_{L,R}= ( L_{L,R}^{0}, L_{L,R}^{-})^T$, and $H_{d}= (H_{d}^{+}, H_{d}^{0})^T$.
When the neutral components of Higgs doublets develop vacuum expectation values, $\left< H_u^0 \right> = v_u$ and $\left< H_d^0 \right> = v_d$, with $\sqrt{v_u^2 + v_d^2} = v = 174$ GeV, we obtain the charged lepton mass matrix
\begin{equation}
	(\bar{\mu}_{L}, \bar{L}_{L}^{-}, \bar{E}_{L})\begin{pmatrix}y_{\mu}v_{d}&0 &\lambda_{E}v_{d}\\\lambda_{L}v_{d}&M_{L}&\lambda v_d\\0&\bar{\lambda}v_{d}& M_{E}\end{pmatrix}\begin{pmatrix}\mu_{R}\\ L_{R}^{-}\\ E_{R}\end{pmatrix}.
\end{equation}
Diagonalizing this matrix we indentify two new mass eigenstates, $e_4$ and $e_5$. The mixing of the 2nd generation of leptons with vectorlike leptons  results in modified couplings of the muon to $W$, $Z$, and $h$,  and new couplings between the muon and heavy leptons. These couplings can be found in Refs.~\cite{Dermisek:2013gta, Dermisek:2015oja} and  Ref.~\cite{Dermisek:2021ajd} and  useful approximations can be obtained from analogous formulas in the quark sector~\cite{Dermisek:2019vkc}, summarized in Ref.~\cite{Dermisek:2021ajd}.

 The total contribution of gauge and Higgs bosons and new leptons to $(g-2)_{\mu}$ can be approximated by~\cite{Dermisek:2020cod, Dermisek:2021ajd}
\begin{equation}
\Delta a_{\mu}  \simeq  - \frac{1+\tan^2 \beta}{16\pi^{2}}  \frac{m_\mu m_\mu^{LE}}{v^2} , 
\label{eq:dela_2HDM}
\end{equation}
 where, 
\begin{equation}
 m_\mu^{LE} \equiv \frac{\lambda_{L} \bar{\lambda} \lambda_{E}}{M_{L}M_{E}} v_d^3.
\end{equation} 
 This approximate formula assumes one scale of new physics $M_{L,E} \simeq m_{H,A,H^\pm}$, however the results are almost unchanged for 
 smaller Higgs masses compared to masses of new leptons. The contributions of SM gauge and Higgs bosons and new leptons sum up to 1 in the numerator of Eq.~(\ref{eq:dela_2HDM}), while the contribution of heavy Higgses and new leptons sum up to $\tan^2\beta$.
This $\tan^2 \beta$ enhancement  is important because, at the same time, these contributions are also enhanced by $\tan^2 \beta$ compared to modifications of the muon couplings to $W$, $Z$ and $h$. Thus increasing the fraction of heavy Higgs contributions to $(g-2)_{\mu}$ decreases the impact of precision EW constraints. This allows to explain $(g-2)_{\mu}$ with very heavy Higgs bosons and new leptons while satisfying precision EW constraints, or with lighter Higgs bosons and new leptons  and almost SM-like muon couplings.

\begin{figure}[t]
\includegraphics[scale=0.25]{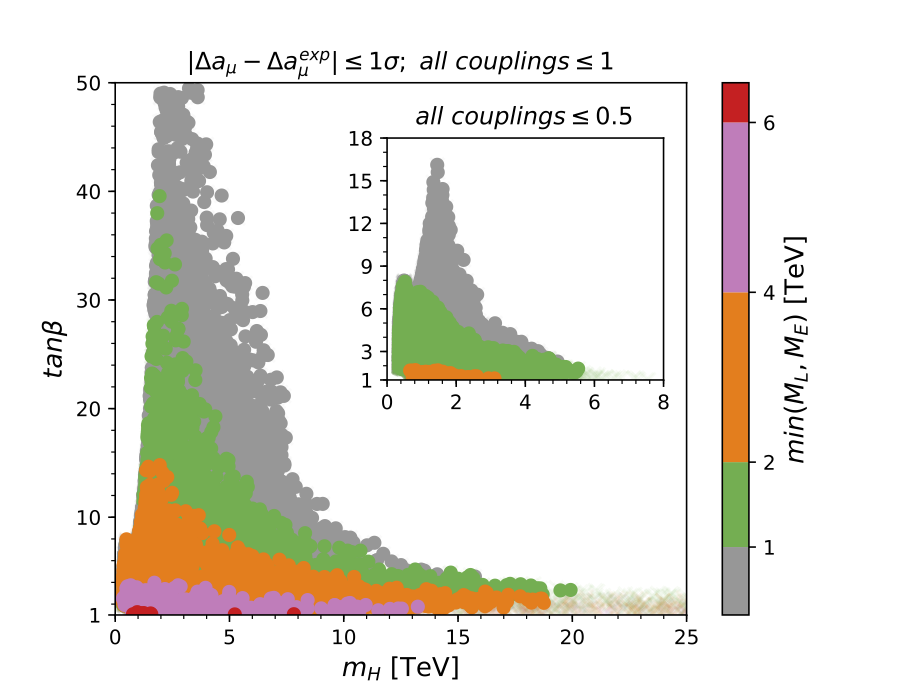} 
\caption{Scenarios satisfying $\Delta a_{\mu}^{\exp}$ within $1\sigma$ in the $m_H - \tan \beta$ plane for Yukawa couplings not exceeding 1 (or 0.5 in the inset). Colors indicate masses of the lightest new lepton. Lightly shaded crosses without filled circles correspond to scenarios with heavy Higgs bosons contributing less than 50\% to $\Delta a_{\mu}$. This plot is from Ref.~\cite{Dermisek:2020cod}.}
\label{fig:MLE}
\end{figure}

In Fig.~\ref{fig:MLE} we plot the scenarios satisfying $\Delta a_{\mu}^{\exp}$ within $1\sigma$ in the $m_H - \tan \beta$ plane limiting the size of any Yukawa coupling by 1 (or 0.5 in the inset). To satisfy constraints from direct searches we require $M_L > 800$ GeV, $M_E > 200$ GeV~\cite{Sirunyan:2019ofn, Aad:2020fzq, Sirunyan:2018mtv}.
Note however that limits depend on the assumed branching ratios of new leptons to $W$, $Z$ and $h$~\cite{Dermisek:2014qca}. In our model an arbitrary pattern of branching ratios can be obtained~\cite{Dermisek:2015hue}. We further assume $m_H = m_A = m_{H^\pm}$ for simplicity, and impose limits on  $H(A) \to \tau^+  \tau^-$~\cite{Aad:2020zxo} and  $H^+ \to t\bar b$~\cite{ATLAS:2020jqj}. With these limits the indirect constraints from flavor physics~\cite{Haller:2018nnx} are satisfied. Constraints from $h \to \mu\mu$~\cite{Aad:2020xfq}, 
the muon lifetime, the $W$ partial width, $Z$-pole observables, and from oblique corrections, summarized in Ref.~\cite{Zyla:2020zbs}, are also imposed.

From Fig.~\ref{fig:MLE} we see that $\Delta a_{\mu}^{\exp}$ can be explained even with 6.5 TeV new leptons or 20 TeV Higgs bosons (with heavy Higgs bosons still providing more than 50\% of the new contribution to  $\Delta a_{\mu}^{\exp}$). Assuming larger Yukawa couplings,  $\sqrt{4\pi}$, the range of masses able to explain  $\Delta a_{\mu}^{\exp}$  extends to  45 TeV for new leptons and 185 TeV for new Higgs bosons~\cite{Dermisek:2020cod}.

\begin{figure}[t]
\includegraphics[scale=0.25]{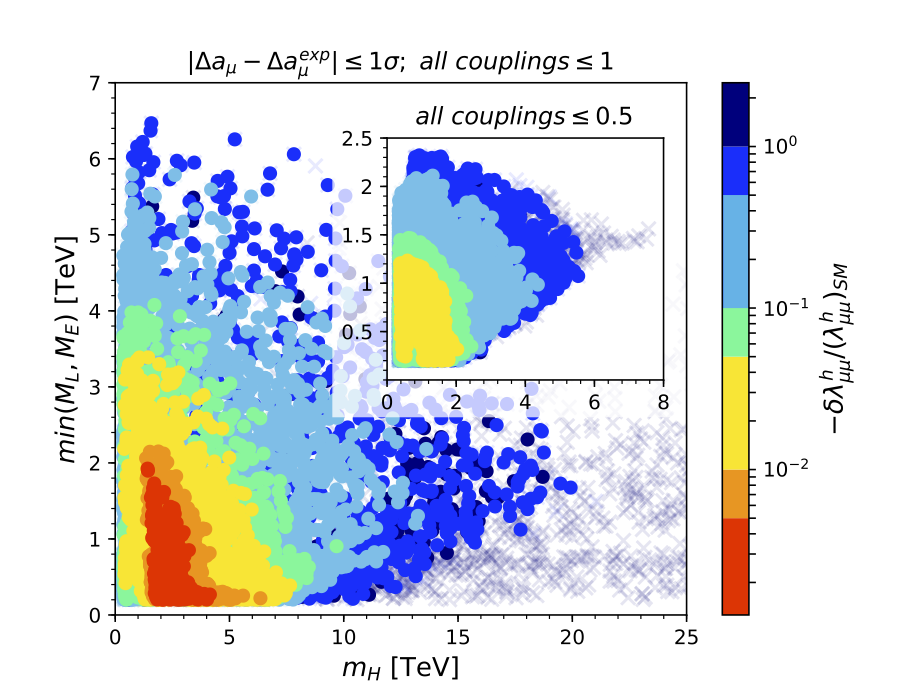}
\caption{The same scenarios as in Fig.~\ref{fig:MLE} in the $m_H - \min(M_L,M_E)$ plane with colors indicating $-\delta \lambda^h_{\mu \mu}/(\lambda^h_{\mu \mu})_{SM}$. This plot is from Ref.~\cite{Dermisek:2020cod}.}
\label{fig:hmumu}
\end{figure}

\begin{figure}[t]
\includegraphics[scale=0.25]{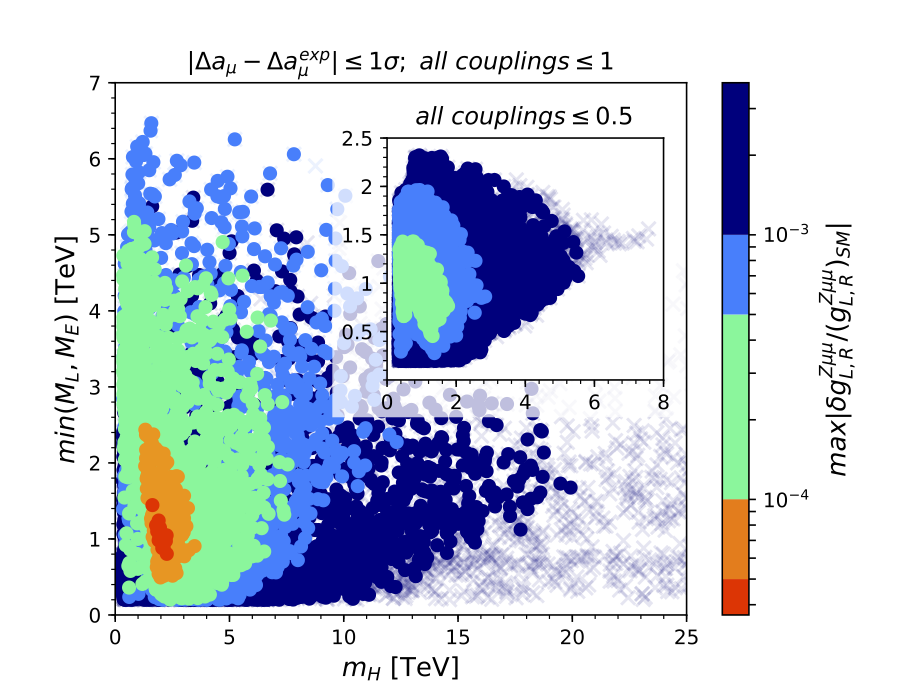}
\caption{The same scenarios as in Fig.~\ref{fig:MLE}  in the $m_H - \min(M_L,M_E)$ plane with colors indicating the largest relative modification of muon couplings to $Z$. This plot is from Ref.~\cite{Dermisek:2020cod}.}
\label{fig:ZWmu}
\end{figure}

In Figs.~\ref{fig:hmumu} and~\ref{fig:ZWmu} we plot the same scenarios in the $m_H - \min(M_L,M_E)$ plane with colors indicating corresponding modification of the muon Yukawa coupling, $-\delta \lambda^h_{\mu \mu}/(\lambda^h_{\mu \mu})_{SM}$, and the the largest modification of muon couplings to the $Z$ boson. Corrections to muon couplings are the largest in the region with small $\tan \beta$ and the heaviest spectrum of new  leptons and Higgs bosons that can explain $\Delta a^{exp}_\mu$. With increasing $ \tan \beta$ the spectrum  has to be lighter, and corresponding corrections to muon couplings decrease by $ \tan^2 \beta$. We see the muon Yukawa coupling and gauge couplings do not have to be affected at more than $10^{-3}$ and $3 \times 10^{-5}$ levels. Assuming larger Yukawa couplings,  $\sqrt{4\pi}$, these ranges decrease to $5 \times 10^{-4}$ and $5 \times 10^{-6}$ levels, respectively~\cite{Dermisek:2020cod}.

Enlarging the model to include SM singlets, $N_{L,R}$, and corresponding Yukawa couplings, to allow mixing in the neutral lepton sector, does not change the results significantly~\cite{Dermisek:2021ajd}. 

The results for the more economical SM with vectorlike leptons can be obtained by replacing: $H_d \to H$, $v_d \to v$, in the formulas above and removing the contribution  of heavy Higgses proportional to $\tan^2\beta$ in Eq.~(\ref{eq:dela_2HDM}). The results are comparable to the results of the 2HDM at small $\tan \beta$.  

Similarly results for extensions of the SM with vectorlike leptons and new scalars that do not participate in electroweak symmetry breaking are also comparable to the results of the 2HDM. In this case, because of no mass mixing of vectorlike leptons with the muon, the impact on muon couplings is very small (reduced to loop corrections), resembling 2HDM at large $\tan\beta$.

\section{Related observables}\label{other}

\begin{figure}[t]
\includegraphics[scale=0.29]{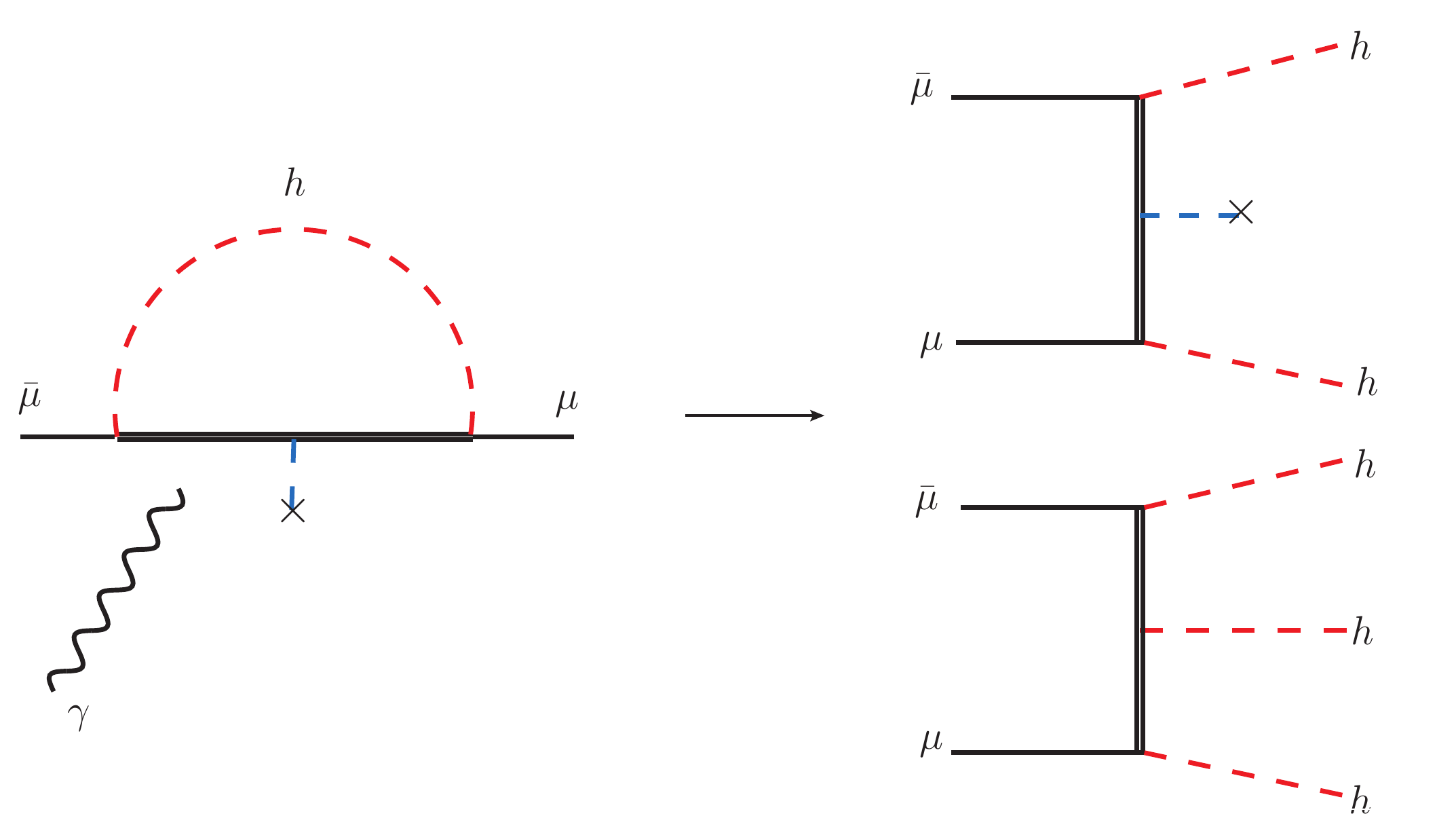}
\caption{Mass enhanced Higgs boson contribution to $(g-2)_{\mu}$ is related to $\mu^+\mu^- \to hh$ (by removing the photon and cutting the loop) and  $\mu^+\mu^- \to hhh$ (by also replacing the vev with $h$). This figure is from Ref.~\cite{Dermisek:2021mhi}.}
\label{fig:g-2_multi-higgs}
\end{figure}

In the SM with vectorlike leptons  there are more observables that are tightly related to  $(g-2)_{\mu}$, see Fig.~\ref{fig:g-2_multi-higgs}. At energy scales much below $M_{L,E} $, the effective Lagrangian:
\begin{equation}
 \mathcal{L}\supset - y_{\mu}\bar{l}_L\mu_{R}H - \frac{\lambda_L \bar \lambda \lambda_E}{M_L M_E}\bar{l}_L\mu_{R}H H^\dagger H + h.c..
 \label{eq:eff_lagrangian}
\end{equation}
is completely fixed by the muon mass mass and muon g-2:
\begin{equation}
m_\mu = y_\mu v + m_\mu^{LE} , \quad \quad \quad \Delta a_{\mu} 
= - \frac{1}{16\pi^{2}}  \frac{m_\mu m_\mu^{LE}}{v^2},
\label{eq:m_mu-a_mu}
\end{equation}
where
\begin{equation}
m_\mu^{LE} \equiv \frac{\lambda_{L} \bar{\lambda} \lambda_{E}}{M_{L}M_{E}} v^3  
\label{eq:m^LE}
\end{equation}
is the contribution to  muon mass from dimension-6 operator. Thus, predictions for all other observables resulting from the effective Lagrangian are unique.
The resulting interactions of the muon with the SM Higgs boson are given by~\cite{Dermisek:2021mhi}:
\begin{equation}
\mathcal{L}\supset - \frac{1}{\sqrt{2}} \, \lambda^h_{\mu\mu}\, \bar{\mu}\mu h - \frac{1}{2} \, \lambda^{hh}_{\mu\mu}\, \bar{\mu}\mu h^2 - \frac{1}{3!}\, \lambda^{hhh}_{\mu\mu}\, \bar{\mu}\mu h^3,
\label{eq:lagrangian_h}	
\end{equation}
where $\mu$ is the Dirac spinor containing $\mu_{L,R}$, and
\begin{eqnarray}
\lambda^h_{\mu \mu} &=& y_\mu + 3 m_\mu^{LE}/v = ( m_\mu +2 m_\mu^{LE})/v \label{eq:lambda_h},\\
\lambda^{hh}_{\mu \mu} &=&  3 \; m_\mu^{LE}/v^2 \label{eq:lambda_hh},\\
\lambda^{hhh}_{\mu \mu} &=& \frac{3}{\sqrt{2}} \;m_\mu^{LE}/v^3 \label{eq:lambda_hhh},
\end{eqnarray}
are predicted without a free parameter.

Modification of the muon Yukawa coupling leads to an altered rate for $h \to \mu\mu$. Defining
\begin{equation}
R_{h\to \mu^+\mu^-} \equiv  \frac{BR(h \to \mu^+\mu^-)}{BR(h \to \mu^+\mu^-)_{SM}} = \left( 1+ 2\frac{m_\mu^{LE}}{m_\mu}\right)^2
\label{eq:R}
\end{equation}
we can easily find that $m_\mu^{LE}/m_\mu = -1.07 \pm 0.25$, required for the explanation of $(g-2)_{\mu}$ within  1$\sigma$,  translates into 
$
R_{h\to \mu^+\mu^-} = 1.32^{+1.40}_{-0.90}.
$
Note that the current upper limit is 2.2~\cite{Aad:2020xfq}.

\begin{figure}[t]
\includegraphics[scale=0.3]{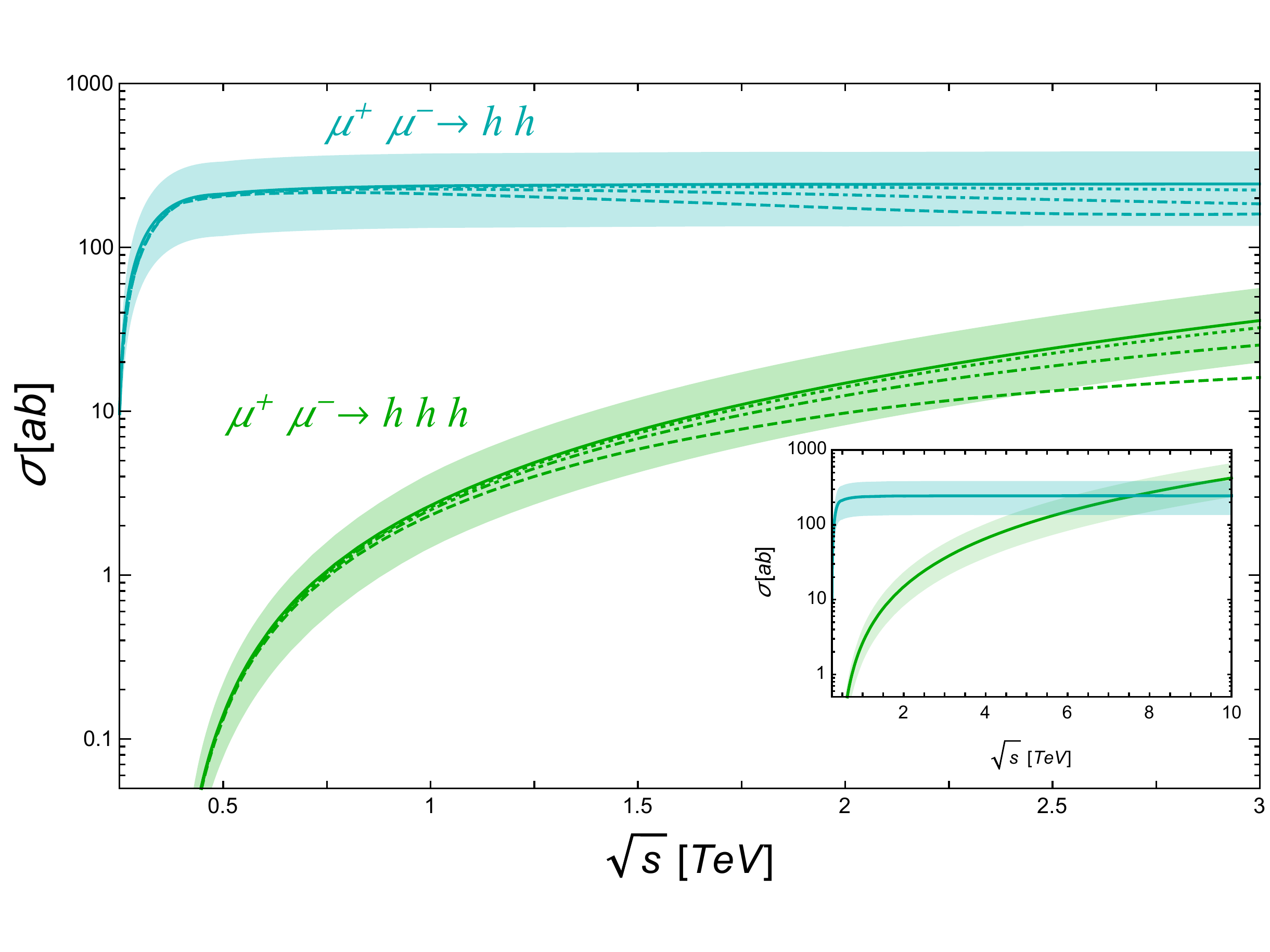}
\caption{Total cross sections for $\mu^+ \mu^- \to hh$ and $\mu^+ \mu^- \to hhh$ as functions of $\sqrt{s}$ corresponding to the  central value of $\Delta a_\mu$ (solid lines) and its  one sigma range (shaded regions) from the effective lagrangian. This plot is from Ref.~\cite{Dermisek:2021mhi}.
}
\label{fig:cross_sections_SM}
\end{figure}

The $\lambda^{hh}_{\mu \mu}$ and $\lambda^{hhh}_{\mu \mu}  $ couplings lead to di-Higgs and tri-Higgs signals at a muon collider. The total cross sections for these processes 
are plotted as functions of $\sqrt{s}$ in Fig.~\ref{fig:cross_sections_SM} for $m_\mu^{LE}$ fixed by 
$\Delta a_\mu$. 
We see that, for example, a muon collider at $\sqrt{s} = 1$ TeV with 0.2 ab$^{-1}$ of integrated luminosity  could see about 50 di-Higgs events, or a  $\sqrt{s} = 3$ TeV collider with 1 ab$^{-1}$ of integrated luminosity  is expected to see about 30 tri-Higgs events (in addition to about 240 di-Higgs events).  SM backgrounds for both these processes are negligible.

In 2HDM the predictions for Higgs related observables discussed above are controlled by $\tan\beta$, and similar new signatures  involving production of heavy Higgses appear~\cite{Dermisek:2021mhi}. In addition to observables directly tight to $(g-2)_{\mu}$, there are other interesting signatures of heavy Higgses and vectorlike leptons, for example flavor violating decays of heavy neutral and charged Higgs bosons through a new lepton~\cite{Dermisek:2015oja, Dermisek:2015hue, Dermisek:2016via}; or decays of new leptons through heavy Higgs bosons (in analogy to  decays  of new quarks~\cite{Dermisek:2019vkc, Dermisek:2020gbr, Dermisek:2021zjd}).

\section{Conclusions}\label{conclusions}

The standard model or two Higgs doublet models with vectorlike leptons can explain $\Delta a^{exp}_\mu$ with multi-TeV new particles and couplings of order 1. The SM case is the most economical and, at the same time, most constrained by precision EW data. It has the most unique predictions that can be searcher for at the LHC ($h\to \mu^+\mu^-$) and at the ILC or FCC-ee (modification of muon gauge couplings). The di-Higgs and tri-Higgs signals at a muon collider are especially striking and can be tested even at very low $\sqrt{s}$.

The 2HDM with vectrolike leptons features $\tan^2\beta$ enhanced contributions of heavy Higgses to $\Delta a_\mu$ compared to contributions of gauge bosons and the SM Higgs boson. This model interpolates between the SM and models with new scalar mediators not participating in EWSB. Because of heavy spectrum of new particles and potentially tiny impact on precision EW observables, testing this explanation of $\Delta a^{exp}_\mu$ would require both the precision of FCC-ee and the energy reach of FCC-hh.

\acknowledgments
I would like to thank K. Hermanek, E. Lunghi, N. McGinnis, and S. Shin for collaborating on projects related to this talk. This work was supported in part by the U.S. Department of Energy under Award No. {DE}-SC0010120.


\end{document}